\documentclass[article,11pt]{article}
\pdfoutput=1
\usepackage{jheppub}

\usepackage{array}
\usepackage{amssymb}
\usepackage{amsbsy}
\usepackage{amsfonts}
\usepackage{amsmath}
\usepackage{epsfig}
\usepackage{color}
\usepackage{graphicx}
\usepackage{epstopdf}
\usepackage{float}
\usepackage{setspace}
\usepackage{amstext}    
\usepackage{wasysym}
\usepackage{array}
\usepackage{hyperref}
\usepackage{ulem}

\newcommand{\blue}[1]{{\color{blue} #1}}

\title{\blue{Limits on Neutrinophilic Two-Higgs-Doublet Models 
from Flavor Physics}}

\author[a]{Enrico~Bertuzzo,} 
\author[a]{Yuber~F.~Perez~G.,}
\author[a,b]{Olcyr~Sumensari,}
\author[a]{Renata~Zukanovich~Funchal}

\emailAdd{bertuzzo@if.usp.br}
\emailAdd{yfperezg@if.usp.br}
\emailAdd{olcyr.sumensari@usp.br}
\emailAdd{zukanov@if.usp.br}

\affiliation[a]{Departamento de F\'{\i}sica Matem\'atica, Instituto de F\'{\i}sica\\
 Universidade de S\~ao Paulo, C.\ P.\ 66.318, 05315-970 S\~ao Paulo, Brazil}
\affiliation[b]{Laboratoire de Physique Th\'eorique (B\^at. 210) \\
CNRS (UMR 8627), Universit\'e Paris-Sud, Universit\'e Paris-Saclay, 91405 Orsay, France}

\abstract{We derive stringent limits on neutrinophilic two-Higgs-doublet 
models from low-energy observables after the discovery of the Higgs boson and of the mixing angle
$\theta_{13}$. These decays can constrain the plane spanned by $m_{H^\pm}$, the mass
of the new charged Higgs, and $v_2$, the vacuum expectation
value of the new neutrinophilic scalar doublet.
Lepton flavor conserving decays are not able to set meaningful bounds, since they depend strongly on the unknown neutrino absolute mass scale. On the other hand, 
loop induced lepton flavor violating decays, such as $\mu \to e \gamma$,
$\mu \to 3 e $ or $\mu \to e$ in nuclei are currently responsable for the 
best limits today.  If $v_2 \lesssim 1 \, (0.1)$ eV we get 
$m_{H^\pm} \gtrsim 250 \, (2500)$ GeV at 90\% CL.
In the foreseen future these limits can improve by at least a factor of 100.}

\begin{document}
\maketitle

\section{Introduction}
%
%
 The discovery of neutrino oscillations leaves basically no doubts on
 the fact that neutrinos are massive particles. Still, whether neutrinos are 
 Dirac or Majorana particles is  an open question. On a general ground, new 
 physics in the neutral and charged lepton sectors are expected to be connected.

 As it is well known, the easiest way to generate neutrino masses is via the addition of at least two
 right-handed (RH) neutrinos to the Standard Model (SM) particle content, with a Yukawa interaction
 ${\cal L} \supset H \overline{\ell}_L Y_\nu
 \nu_R$. Since the $\nu_R$ are necessarily
 gauge singlets, no gauge symmetry can forbid a Majorana mass term
 ${\cal L} \supset M_R \nu_R \nu_R$. Light neutrinos are a natural 
 outcome in a seesaw scenario~\cite{seesaw}, where $M_R$ is assumed to be very large.
 Once the RH neutrinos are integrated out, the Weinberg operator $(\ell
 H)^2/\Lambda$ is generated~\cite{Weinberg:1979sa}, producing Majorana
 neutrinos. In addition, heavy RH neutrinos may also generate the right amount of matter-antimatter asymmetry
 through leptogenesis~\cite{Fukugita:1986hr}. 
 
 On the contrary, if we insist on light RH neutrinos ({\it i.e.} small
 $M_R$), small neutrino masses require $Y_\nu \sim 10^{-(12\div 13)}$,
 roughly $7$ orders of magnitude smaller than the electron
 Yukawa coupling. Moreover, the presence of the RH Majorana mass makes the
 neutrinos pseudo-Dirac particles, rather than Dirac. In this case,
 baryogenesis is still possible through neutrinogenesis~\cite{Dick:1999je}, but  the
 stringent cosmological limits on the number of relativistic degrees of freedom
 (RDOF)  obtained by Planck~\cite{Ade:2015xua} are a potential shortcoming. The easiest way out is to impose 
 the light $\nu_R$ to not  contribute at all to the RDOF, {\it i.e.} to decouple from the plasma well before
 Big Bang Nucleosynthesis.
 
 The tiny Yukawa couplings in the pseudo-Dirac case are of course a
 possibility, as they may find an explanation in a theory of flavor
 (just like any theory of flavor would explain the $5$ orders of
 magnitude difference between the top and the electron Yukawa
 couplings). However, to get pure Dirac states, an additional symmetry
 must be imposed to forbid the RH Majorana mass. To avoid charging
 also some SM state under the same symmetry, this is most
 easily realized decoupling neutrino masses from the SM Higgs
 doublet. A second scalar, charged under the additional
 symmetry, can thus be introduced, whose vacuum expectation value (vev) is responsible for neutrino
 masses~\cite{Gabriel:2006ns}. If the vev happens to be in the eV range, we obtain the correct order of magnitude for the neutrino
 masses with ${\cal{O}}(1)$ Yukawa couplings. Models with such a
 {\it neutrinophilic} doublet and pure Dirac neutrinos were studied in
 some detail in~\cite{Davidson:2009ha}.  Recently, after the 
discovery of the Higgs boson by the LHC experiments, 
these minimal Neutrinophilic Two-Higgs-Doublet Models ($\nu$2HDM) were revisited 
and strong bounds were imposed on their scalar spectrum~\cite{Machado:2015sha}.

We focus here on additional experimental consequences from  
charged lepton flavor physics on $\nu$2HDM, in particular, 
after the last neutrino mixing angle, $\theta_{13}$, was 
precisely measured by the reactor experiments Daya Bay~\cite{An:2015rpe} and
Double Chooz~\cite{Abe:2014bwa}. Note that non-standard interactions in the neutrino 
sector only involve RH neutrinos in these models, so they are not expected to affect neutrino oscillations.

In Sec.~\ref{sec:model} we describe the main features of 
$\nu$2HDM that are important to understand how they modify lepton 
flavor physics.
In Sec.~\ref{sec:exp_constr} we calculate the most constraining 
processes involving charged leptons that are affected by  $\nu$2HDM: 
tree-level flavor conserving leptonic decays, loop induced  flavor 
violating leptonic decays, tree-level flavor violating  
$Z$ and Higgs decays. We derive current bounds on the model 
parameters by using the most precise experimental data available 
today and estimate the improved sensitivity of future experiments to 
these parameters. In Sec.~\ref{sec:conc} we present  our final conclusions.

\section{A brief description of the model}\label{sec:model}

Let us now describe the class of models we are interested in. We
extend the SM particle content to include a second Higgs doublet
$H_2$, as well as three RH neutrinos $\nu_R$. $H_2$ has the same gauge
quantum numbers as the ordinary Higgs $H_1$, while the RH neutrinos
$\nu_R$ are gauge singlets. The interactions we are interested in are
given by
\begin{equation}\label{eq:Yukawa_lagr}
 -{\cal L}_{\rm Yuk} = \overline{e}_R Y_E H_1 \ell_L + \overline{\nu}_R Y_N \widetilde{H}_2 \ell_L\, +\mathrm{h.c.},
\end{equation}
where $\widetilde{H}_2=i \sigma_2 H_2^\ast$ is the conjugate of $H_2$.
The previous lagrangian can be easily obtained requiring $H_2$ and
$\nu_R^i$ to have the same charge under an additional global
$U(1)_X$~\cite{Davidson:2009ha}, or imposing a ${\mathbb Z}_2$
symmetry~\cite{Gabriel:2006ns}. In the following we will not be
concerned with the specific realization from which
Eq.~(\ref{eq:Yukawa_lagr}) is obtained, focusing only on its
consequences on flavor processes.

Using the PMSN matrix $U$ to express the gauge eigenstates $\nu_\alpha$ in terms of 
the mass eigenstates $\nu_i$, $\nu_\alpha = U_{\alpha i} \nu_i$, from Eq.~(\ref{eq:Yukawa_lagr}) we get
%
\begin{equation}
 \displaystyle{-{\cal L}_{\rm Yuk} = H_1^0 \overline{e} \hat{Y}_E P_L e + H_2^0 \overline{\nu}_i \hat{Y}_N^i P_L \nu_i - H_2^+ \overline{\nu}_i (\hat{Y}_N U^\dag)_{i\beta} P_L e_\beta +\mathrm{h.c.} \, ,}
\end{equation}
where $\hat{Y}_{E,N}$ are diagonal matrices of Yukawa couplings. Assuming $H_2$ to acquire a vev 
$v_2 \lesssim {\cal O}({\rm eV}) \ll v_1 \simeq 246$ GeV, we get that ${\cal O}(1)$ Yukawa 
couplings in the neutrino sector are possible. Moreover, since $v_2 \ll v_1$,  we can identify $H_1$ 
with the SM Higgs doublet, while the second doublet can be written in terms of the additional scalar mass eigenstates, $H$, $A$ and $H^+$, as $H_2 \simeq (-H^+, \frac{H+iA}{\sqrt{2}})^T$. Let us stress that if an exact $U(1)_X$ symmetry is present also in
the scalar potential, $A$ is a strictly massless Goldstone boson. This calls for an explicit symmetry breaking to make the model phenomenologically viable, which in~\cite{Davidson:2009ha} is given by a soft term in the potential.\footnote{A massless neutrinophilic scalar would be in conflict with several constraints, such as stellar cooling \cite{Zhou:2011rc} and electroweak precision tests \cite{Machado:2015sha}. In addition, compatibility with the total number of relativistic degrees of freedom measured by Planck~\cite{Ade:2015xua} would impose very strong constraints on its parameter space (see Appendix~\ref{appendix}).}

Given its importance for the following sections, let us rewrite the coupling involving the charged Higgs in the neutrino flavor basis:
\begin{equation}\label{eq:Leff}
 -{\cal L}_{\rm charged} = \sqrt{2}\frac{U_{\alpha i} m_{\nu_i} U^*_{\beta i}}{v_2} H_2^+ \overline{\nu}_{\alpha} P_L e_{\beta}\ +\mathrm{h.c.}, ~~~\alpha,\beta = e, \, \mu, \, \tau \, .
\end{equation}
It is clear that apart from the $v_2$ dependence, the coupling is completely fixed in terms of (known) neutrino parameters. Integrating out the massive charged Higgs boson, we obtain
%
\begin{equation}\label{eq:eff_L}
  {\displaystyle -{\cal L}_{\mathrm{eff}} = \frac{1}{m^2_{H^\pm}}\frac{\langle m_{\alpha\beta} \rangle}{v_2} \frac{\langle m_{\rho\sigma} \rangle}{v_2} \left( \overline{\nu}_\alpha \gamma^\mu P_R \nu_\sigma \right) \left(\overline{e}_\rho \gamma_\mu P_L e_\beta \right) + \dots \, ,}
\end{equation}
where $\langle m_{\alpha\beta} \rangle = U_{\alpha i} m_{\nu_i} U^*_{\beta i}$ and 
only the non-standard neutrino interaction term is displayed. As already 
anticipated in the introduction, this term only involves RH neutrinos. 
Since Eq.~(\ref{eq:eff_L}) does not have an analog term in the quark sector, 
we do not expect RH neutrinos to be produced at a relevant rate by nuclear 
processes, for instance, in the sun, in such a way that there is no modification of neutrinos 
propagation through matter.

\section{Experimental Constraints from Charged Lepton processes}
\label{sec:exp_constr}

The phenomenology of a neutrinophilic charged Higgs in low-energy
processes is different from the one of a generic 2HDM, mainly because
the couplings with leptons are highly enhanced by a factor $v_1/v_2 \gg 1$,
while the couplings with quarks are highly suppressed by $v_2/v_1 \ll 1$. 
As a first consequence, the $\nu$2HDM easily evades
the limits coming from hadronic observables such as $B\to X_s \gamma$,
mesons mixing and, more importantly, leptonic and semi-leptonic $B$
meson decays~\cite{Lees:2012xj,Deschamps:2009rh,Crivellin:2013wna}. On
the other hand, leptonic observables (normally suppressed by neutrino
masses) may now receive sizable contributions, as we are now going to
show. Our main results are summarized in Fig.~\ref{fig:limits}, where we show 
the current bounds (left panel) and the expected future sensitivities (right panel) on the $(m_{H^\pm}, v_2)$ plane.

\begin{figure}[tb]
 \begin{center}
   \includegraphics[width=\textwidth]{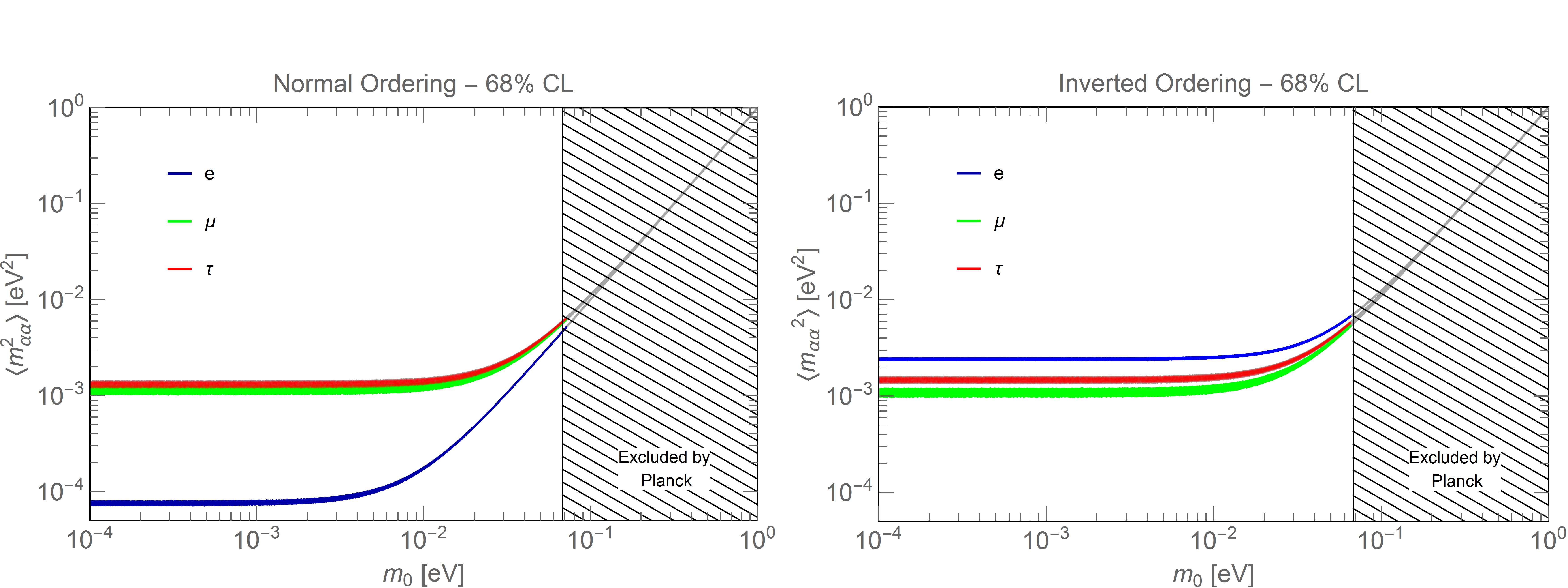}
 \end{center}
 \caption{\label{fig:maa} Values of $\langle m^2_{\alpha\alpha}
   \rangle$ as a function of the lightest neutrino mass $m_0$ obtained
   scanning over the $1 \sigma$ range of the oscillation parameters~\cite{Gonzalez-Garcia:2014bfa}. Blue region: $\alpha =
   e$, green region: $\alpha = \mu$, red region: $\alpha = \tau$.  The left panel refers to the Normal neutrino mass ordering (NO), the right
   panel to the Inverted ordering (IO). The gray points are excluded by Planck's
   limit on the sum of neutrinos masses~\cite{Ade:2015xua,Giusarma:2013pmn}.}
\end{figure}

\subsection{Lepton Flavor Conserving Decays}

Let us look at the tree level $\mu$ and $\tau$ leptonic decays. The
charged scalar contribution to $\ell_\alpha \to \ell_\beta
\bar{\nu}\nu$ induces a violation of lepton flavor universality (LFU),
which can be effectively encoded in the definition of ``flavorful''
gauge couplings $g_\alpha$~\cite{Roney:2007zza}. Experimentally, they
can be measured from the $\tau$ and $\mu$ lifetimes. The total decay width for 
$\ell_\alpha \to \ell_\beta \overline{\nu}\nu$ in the presence of a charged Higgs 
boson can be written as $\Gamma(\ell_\alpha \to \ell_\beta \overline{\nu}\nu) = \Gamma^{\rm SM}(\ell_\alpha \to \ell_\beta \overline{\nu}\nu) (1+ \langle m^2_{\alpha\alpha} \rangle \langle m^2_{\beta\beta} \rangle \rho^2 /8)$~\cite{Fukuyama:2008sz,Davidson:2009ha}, from which
\begin{equation}\label{eq:ratiog}
 \begin{array}{ccl}
  \left(\dfrac{g_\mu}{g_e}\right)^2 &\simeq & 1+ \dfrac{\langle m_{\tau\tau}^2 \rangle (\langle m_{\mu\mu}^2 \rangle-\langle m_{ee}^2 \rangle)}{8 } \rho^2, \\
\left(\dfrac{g_\mu}{g_\tau}\right)^2 &\simeq&1+ \dfrac{\langle m_{ee}^2 \rangle (\langle m_{\mu\mu}^2 \rangle-\langle m_{\tau\tau}^2 \rangle)}{8} \rho^2.\,
 \end{array}
\end{equation}
We have defined $\langle m_{\alpha \beta}^2 \rangle = U_{\alpha i} m_{\nu_i}^2 U^*_{\beta i}$ and $\rho =(G_F m_{H^\pm}^2 v_2^2)^{-1}$.

Although from Eq.~(\ref{eq:ratiog}) it may look like it is possible to
use the experimental results on $g_\mu/g_e$ and $g_\mu/g_\tau$ to
extract bounds on $\rho$, we stress that (i) the experimental data
are well compatible with lepton flavor universality at $1\, \sigma$,\footnote{We are aware of the 2$\sigma$ disagreement 
between the PDG and HFAG fits of $B_\mu=\mathrm{BR}(\tau\to\mu\nu_\tau\bar{\nu}_\mu)$ with respect to the SM \cite{Agashe:2014kda,Amhis:2014hma}. However, we notice that the world average of $B_\mu$, which we are considering here, is in perfect agreement with the SM prediction. The main difference between these two results is the inclusion of the ratio $B_\mu/B_e$ measured precisely by BaBar, which has a slight disagreement of $1.6\sigma$ with the leptonic universality assumption \cite{Aubert:2009qj}.} and (ii) although the differences $\langle m^2_{\alpha\alpha} \rangle - \langle m^2_{\beta\beta}
\rangle$ are independent of the value of the lightest neutrino mass
$m_0$, the individual values of $\langle m^2_{ee} \rangle$ and $\langle m^2_{\tau\tau} \rangle$ depend crucially on $m_0$, as shown in Fig.~\ref{fig:maa}. From this plot, it is clear that the flavorful couplings reach very small values for $m_0^2\ll \Delta m_{ij}^2$, making them always compatible with the currents bounds on LFU \cite{Roney:2007zza}. Since the absolute neutrino mass scale is still unknown, no information can be extracted from these observables.

\subsection{Lepton Flavor Violating Decays}
We will study here loop induced lepton flavor violating (LFV) 
processes that can currently, or in the near future, be constrained by data, 
and the corresponding consequences on the allowed values of the 
$\nu$2HDM parameters. 

\subsubsection{$\ell_\alpha \to\ell_\beta \gamma$}
Let us start with loop induced processes, for which strong
experimental constraints are available, at least in the $\mu \to e
\gamma$ channel. For a generic process $\ell_\alpha \to\ell_\beta
\gamma$, the scalar mediated branching ratio
reads~\cite{Fukuyama:2008sz}
\begin{equation}
\mathrm{BR}(\ell_\alpha\to \ell_\beta\gamma) = \mathrm{BR}(\ell_\alpha\to e\bar{\nu}\nu)\frac{\alpha_\text{EM}}{192 \pi} |\langle m_{\alpha \beta}^2 \rangle|^2 \rho^2 \, .
\end{equation}

The strongest experimental bound on this type of process comes from
the MEG-2 upper limit $\mathrm{BR}(\mu\to e \gamma)<5.7 \times
10^{-13}$~\cite{Adam:2013mnn}, while weaker bounds on the other channels are
obtained by the BaBar Collaboration, $\mathrm{BR}(\tau \to e
\gamma) < 3.3 \times 10^{-8}$ and $\mathrm{BR}(\tau \to \mu \gamma)<
4.4 \times 10^{-8}$~\cite{Aubert:2009ag}. In terms of $\rho$ (defined below Eq.~(\ref{eq:ratiog})), 
we get the $90\%$ C.L. bounds\footnote{These limits have a very loose dependence 
  on the neutrino mass hierarchy. Here, we always present the most pessimistic bound, 
  which for $\mu \to e \gamma$ and $\tau \to \mu \gamma$ is obtained in the NO, 
  while for $\tau \to e \gamma$ the IO is slightly less constraining.}
\begin{equation}
 \begin{array}{ll}
  \rho \lesssim 1.2 \, \mathrm{eV}^{-2} & ~~~~~[\mu \to e \gamma]\, , \\
  \rho \lesssim 730 \, \mathrm{eV}^{-2} & ~~~~~[\tau \to e \gamma]\, , \\
  \rho \lesssim 793 \, \mathrm{eV}^{-2} & ~~~~~[\tau \to \mu \gamma]\, . \\
 \end{array}
\end{equation}
This is the best limit at present on the parameters $v_2$ and $m_{H^\pm}$, and it already 
implies that, insisting on $v_2 \lesssim 1$ eV, we must have $m_H^{\pm}\gtrsim 250$ GeV. 
With the future improvement on the MEG expected sensitivity, $\mathrm{BR}(\mu \to e \gamma) \sim 5
\times 10^{-14}$~\cite{Baldini:2013ke,Iwamoto:2014tfa}, the corresponding bound on $\rho$ 
can be improved by about one order of magnitude to $\rho
\lesssim 0.4 \; \mathrm{eV}^{-2}$.
The limits imposed by the MEG bound on the $(m_{H^\pm}, v_2)$ plane are shown in
Fig.~\ref{fig:limits}, blue line, for the current result (left panel), 
as well as for the expected future sensitivity 
(right panel).

\begin{figure}[tb]
  \begin{center}
    \includegraphics[width=0.44\textwidth]{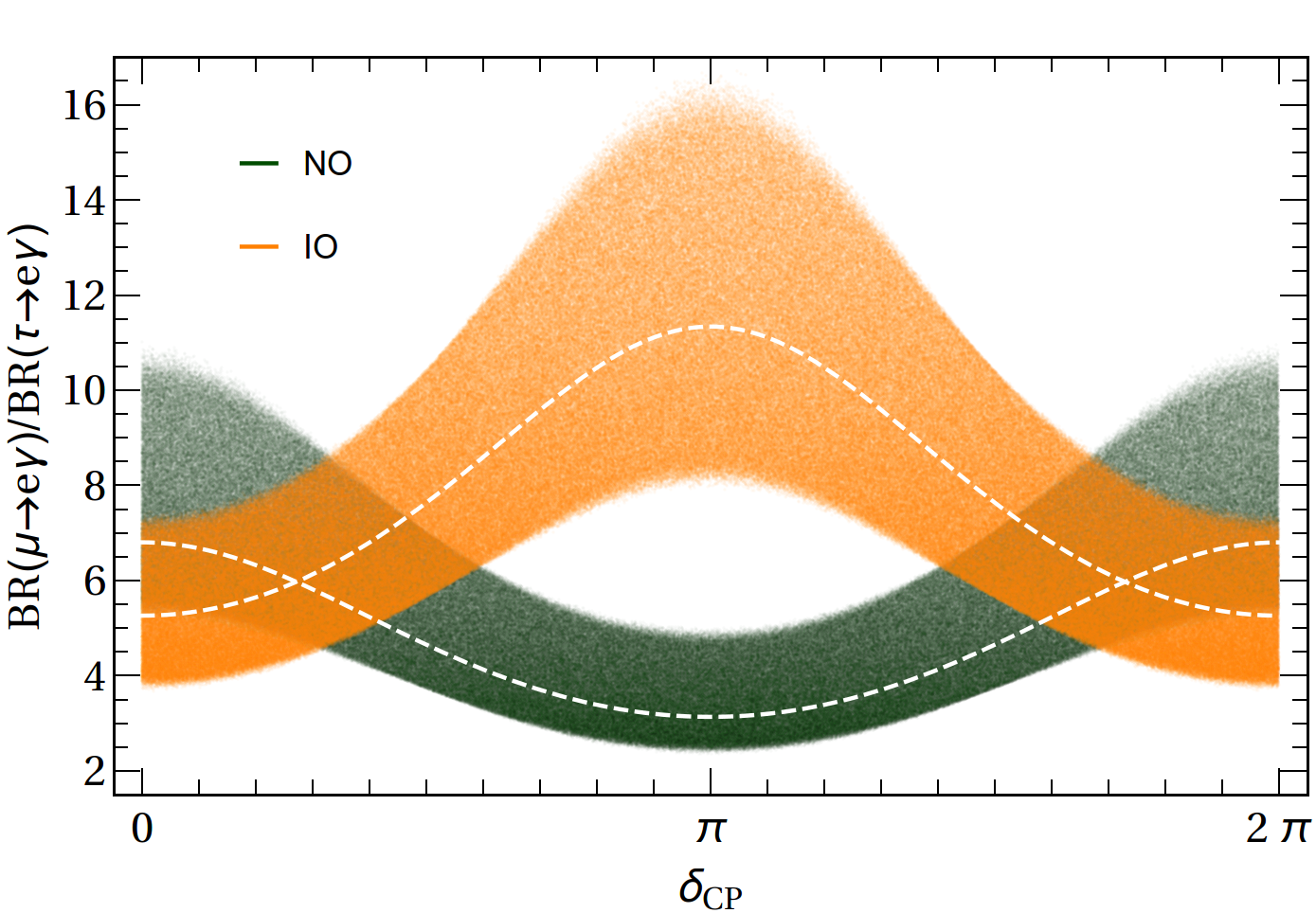} 
    \includegraphics[width=0.455\textwidth]{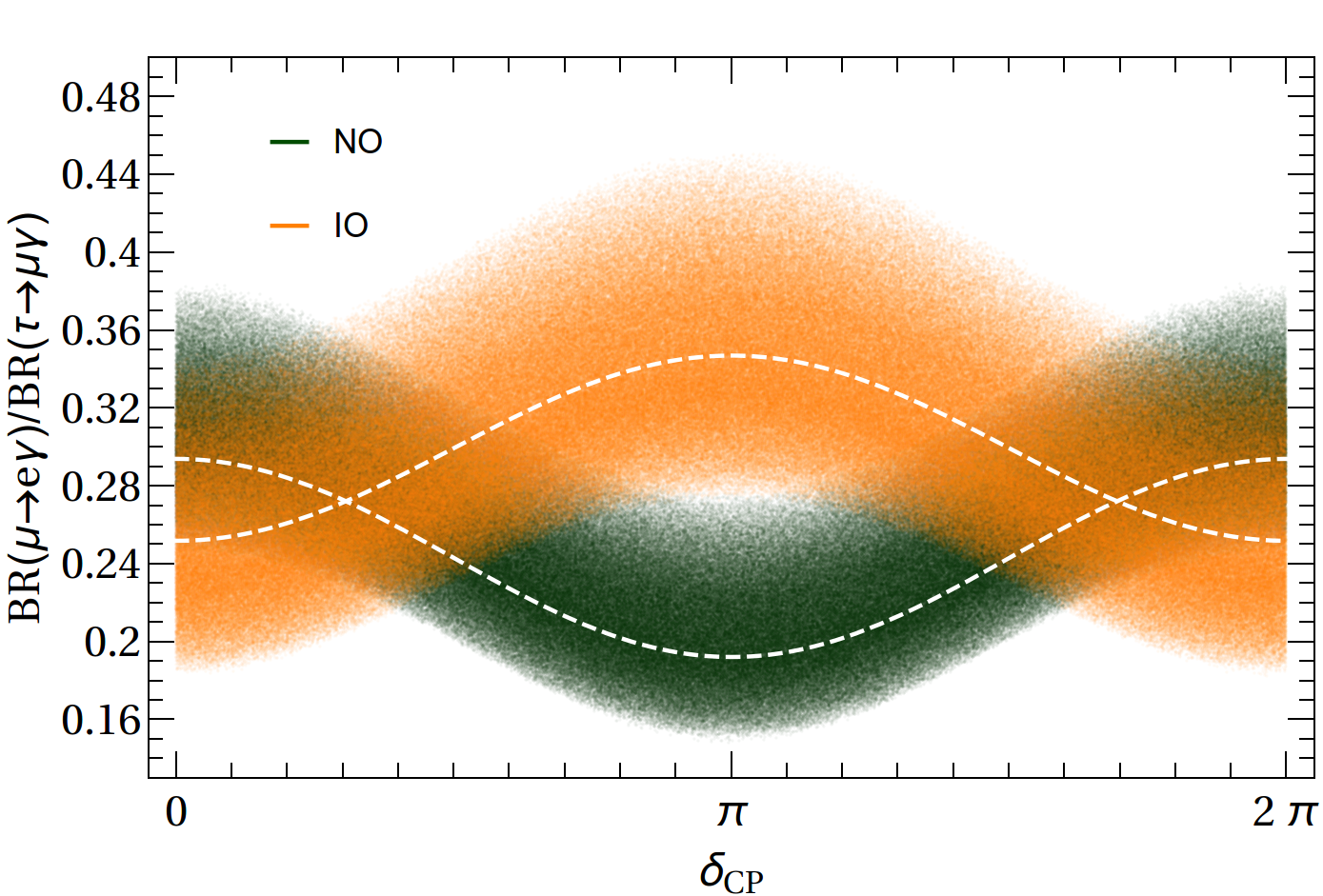} 
  \end{center}
  \vspace{-0.5cm} \caption{\label{fig:cp} 
    Ratios $\mathrm{BR}(\mu \to e \gamma)/\mathrm{BR}(\tau \to e \gamma)$ (right panel) and 
    $\mathrm{BR}(\mu \to e \gamma)/\mathrm{BR}(\tau \to \mu \gamma)$ (left panel)
    as a function of $\delta_{CP}$. The bands were obtained by scanning over the $2\sigma$ range 
    of the oscillation parameters with respect to the central values (white dashed lines), which
    correspond to the best fit \cite{Gonzalez-Garcia:2014bfa}. }
\end{figure}

We would like to point out that in case a positive sign of $\mu \to e \gamma$ is observed in the near future,  
 $\nu$2HDM predicts a relation between $\mathrm{BR}(\mu\to e \gamma)$ and $\mathrm{BR}(\tau\to e \gamma, \mu \gamma)$ 
which is actually sensitive to $\delta_{\rm CP}$. We show in Fig.~\ref{fig:cp} the ratio of these branching 
ratios as a funtion of $\delta_{CP}$. Although it is very unlikely to be able to experimentally probe 
$\mathrm{BR}(\tau\to e\gamma, \mu\gamma)$ down to $10^{-14}$ in the near future, from Fig. \ref{fig:cp}
we see that a limit on $\mathrm{BR}(\mu\to e\gamma)$ would also set a stringent limit on $\mathrm{BR}(\tau\to e\gamma, \mu\gamma)$ independently of $\delta_{CP}$. Moreover, if $\delta_{CP}$ can be measured by neutrino oscillations experiments, this result can be translated 
into a correlation of the different LFV branching ratios.

\subsubsection{$\ell_\alpha \to 3 \, \ell_\beta$}

We now turn our attention to processes involving three charged leptons in the 
final state. These processes can be described by the loop induced 
effective lagrangian
\begin{equation}
\mathcal{L}_{\mathrm{eff}}=  \frac{e \, m_\alpha}{2}A_D\bar{\ell}_\beta \sigma_{\mu\nu}\ell_\alpha F^{\mu\nu}+ e A_{ND} \bar{\ell}_\beta \gamma_\mu P_L \ell_\alpha A^\mu+e^2 B (\bar{\ell}_\alpha \gamma_\mu P_L \ell_\beta) (\bar{\ell}_\beta \gamma_\mu P_L \ell_\beta)+\mathrm{h.c.},
\end{equation}
where $m_{\alpha}$ is the mass of the charged lepton in the initial
state, and $A_{(N)D}$ and $B$ are Wilson coefficients associated to
$\gamma$-penguin diagrams and charged Higgs boxes, respectively.
Neglecting neutrino masses and imposing $m_\beta/m_\alpha\ll 1$, we
derived the following expressions,
\begin{align}
\label{eq:AD}A_D &= \frac{1}{6(4\pi)^2}\frac{1}{m_{H^\pm}^2}\frac{\langle m_{\alpha \beta}^2\rangle}{v_2^2}\, ,\\
\label{eq:AND}A_{ND} &= \frac{1}{9(4\pi)^2}\frac{q^2}{m_{H^\pm}^2}\frac{\langle m_{\alpha \beta}^2\rangle}{v_2^2}\, , \\
\label{eq:B}e^2 B &= -\frac{2}{(4\pi)^2}\frac{\langle m_{\alpha \beta}^2\rangle\langle m_{\beta \beta}^2\rangle}{m_{H^\pm}^2 v_2^4},
\end{align}
where $q^2$ is the photon squared momentum in the penguin-like
diagrams~\footnote{Notice that the dimension four contribution
  vanishes for off-shell photons, as required by gauge
  invariance.}. We neglect the $Z$-penguin diagrams, since they are suppressed by
$m_\beta$ and by the $Z$ boson mass.

Our results for the Wilson coefficients are in full agreement with
those of Ref.~\cite{Toma:2013zsa}, where the impact of the scotogenic
model on $\mu\to 3 e$ was studied. Although these models are
intrinsically different, one can retrieve the $\nu$2HDM loop functions
by replacing the mass of RH neutrinos by the active ones and
by matching the Yukawa lagrangians of the two models. The only
difference between the two calculations is a new
$Z$-penguin diagram with two neutrinos in the loop, which is only
present in the $\nu$2HDM model (see Fig.~\ref{fig:lfvz}). This diagram does not
exist in the scotogenic model, because the $\mathbb{Z}_2$ symmetry
forbids the mixing between the active and sterile neutrinos. However,
this diagram is additionally suppressed by the active neutrino
mass, giving a negligible contribution.

In terms of the coefficients defined above, the branching ratio reads
\begin{align}
\mathrm{BR}(\ell_\alpha\to \ell_\beta\ell_\beta\ell_\beta) = \mathrm{BR}(\ell_\alpha\to e\bar{\nu}\nu) &\frac{3 (4\pi)^2\alpha_{\mathrm{EM}}^2}{8 \, G_F^2} \Bigg{[} \frac{|A_{ND}|^2}{q^4}+|A_D|^2\left(\frac{16}{3}\log\left(\frac{m_\alpha}{m_\beta}\right)-\frac{22}{3} \right) \nonumber \\
&\hspace{-1.5cm}+\frac{1}{6}|B|^2+2\,\mathrm{Re}\left(-2 \frac{ A_{ND}}{q^2} A_D^\ast+\frac{1}{3}\frac{A_{ND}}{q^2}B^\ast-\frac{2}{3}A_D B^\ast \right)\Bigg{]},
\end{align}
where the sub-leading terms in $m_\beta/m_\alpha$ have been
neglected. Notice that the box terms have a different dependence on
$v_2$ and therefore this expression cannot be expressed only in terms
of $\rho$. Moreover, from Eqs.~(\ref{eq:AD}-\ref{eq:B})
it is clear that the box contribution dominates for small
values of $v_2$.

For relatively large values of $v_2$, in the region where the penguin diagrams dominate, 
we can use the current experimental limit, $\rm{BR}(\mu \to e^{-} e^{-} e^{+})< 1 \times 10^{-12}$~\cite{Bellgardt:1987du} 
to directly put the bound $\rho \lesssim 22$ eV$^{-2}$. This is not possible for small $v_2$, 
in the region where the box dominates, since the corresponding Wilson coefficient cannot be 
expressed in terms of $\rho$. The total bound is shown in Fig.~\ref{fig:limits}, green line. 
The current experimental limit (left panel), is stronger than the limit derived from 
$\mu \to e \gamma$ for $v_2 \lesssim 0.01$ eV. The situation will change with the future 
Mu3e experiment, which aims to reach an ultimate sensitivity of 
$\rm{BR}(\mu \to e^{-} e^{-}e^{+}) \sim 1 \times 10^{-16}$~\cite{Blondel:2013ia}. As can be 
seen from Fig.~\ref{fig:limits}, right panel, the future sensitivity on $\mu\to 3e$ is expected to be 
stronger than the one from $\mu \to e \gamma$.

\subsubsection{$\mu\to e$ in nuclei}
\begin{figure}[tb]
 \begin{center}
   \includegraphics[width=0.45\textwidth]{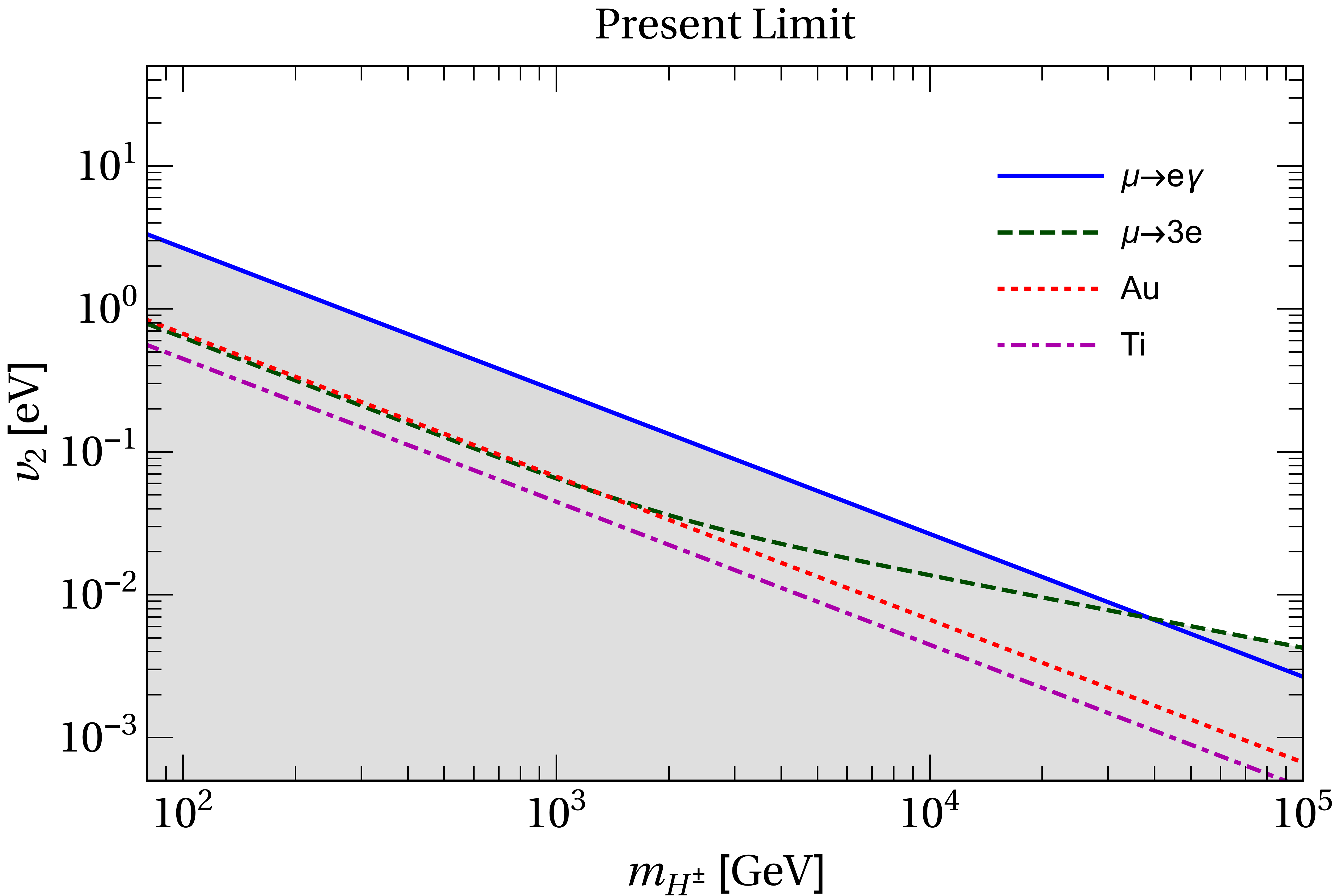}
   \includegraphics[width=0.45\textwidth]{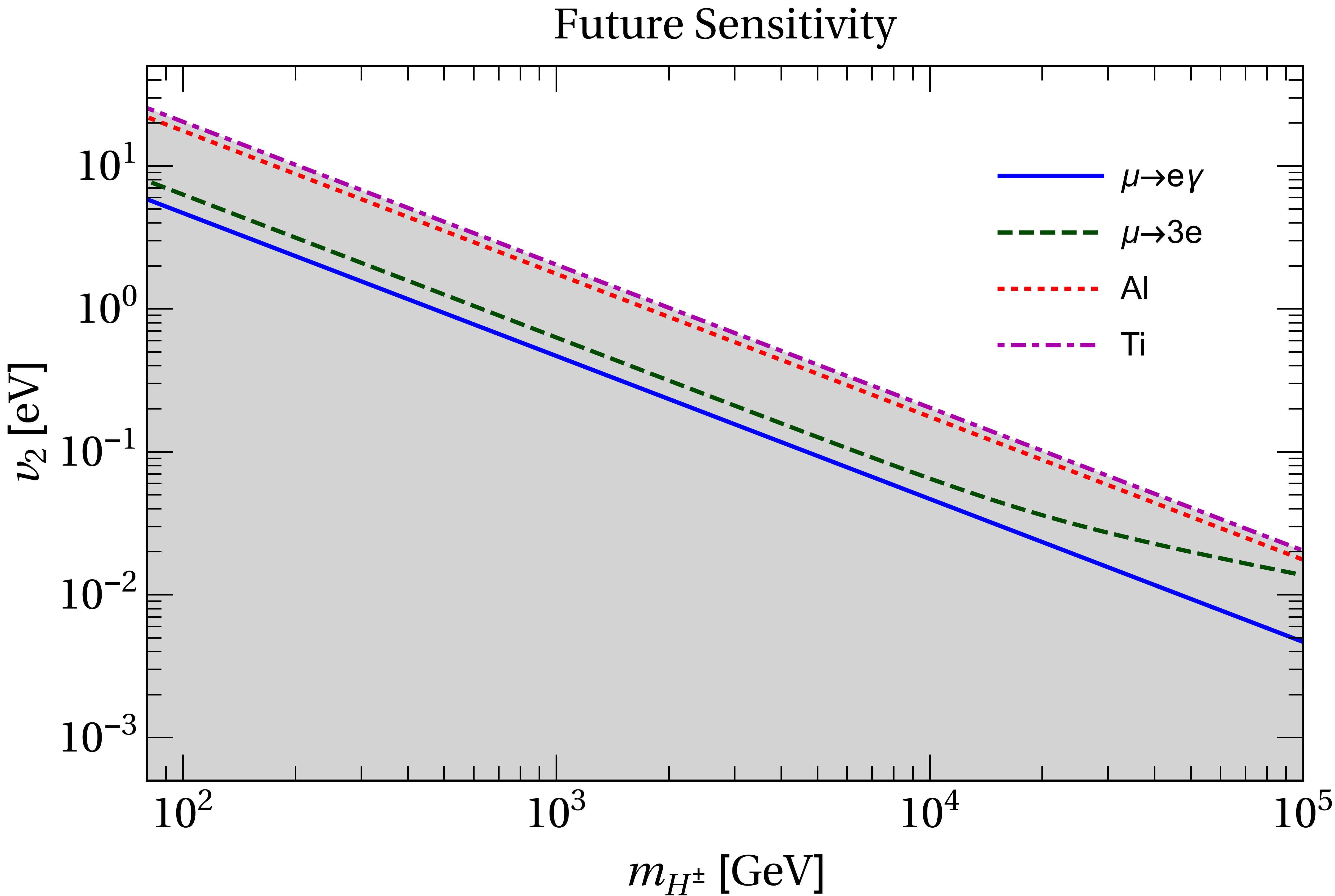}
 \end{center}
 \caption{\label{fig:limits} On the left panel we show the current limits on 
$(m_{H^\pm}, v_2)$ plane coming from $\mu \to e \gamma$, $\mu \to eee$ and $\mu \to e$ 
in nuclei. The regions below these lines are excluded at 90\% CL. 
 On the right panel we show the predicted sensitivities of future experiments. The parameter 
region above each of these lines can be explored by the respective experiment.}
\end{figure}

\begin{table}[tb]
\centering
\begin{tabular}{|c|c|c|c|}\hline
$^A_Z$ Nucleus  & $Z_{\text{eff}}$ & $F_p$ & $\Gamma_{\text{capt}}$ (GeV)\\ \hline\hline
$^{27}_{13}\text{Al}$ & 11.5 & 0.64 & $4.64079 \,\times\,10^{-19}$\\ 
$^{48}_{22}\text{Ti}$ & 17.6 & 0.54 & $1.70422\,\times\, 10^{-18}$\\ 
$^{197}_{\phantom{1}79}\text{Au}$ & 33.5 & 0.16 & $8.59868\,\times\,10^{-18}$\\ \hline
\end{tabular}
\caption{Nuclear parameters used in our analysis, taken from~\cite{Arganda:2007jw}.\label{tab:TabNuclear}}
\end{table}
The $\mu-e$ conversion in nuclei is another LFV process that appears 
in $\nu$2HDM. It is important to note that 
the experimental collaborations have announced great future sensitivities, making this a relevant
bound for different neutrino mass models. In our framework, the dominant contributions are only the $\gamma$-penguins, since the $Z$-penguins are suppressed by the electron and $Z$ boson masses, while box diagrams and scalar penguins are suppressed by the tiny coupling of the neutrinophilic scalars with quarks. Keeping only the dominant contributions, the conversion rate is given by 
\cite{Toma:2013zsa,Kuno:1999jp,Arganda:2007jw}
\begin{align}
  \text{CR}(\mu-e,\text{nucleus})=\frac{p_e\, E_e\, m_\mu^3\, G_F^2\, \alpha_\text{EM}^3\,Z_{\text{eff}}^4\,F_p^2}{8\pi^2\,Z\,\Gamma_{\text{capt}}}&\ \left|(Z+N)g^{(0)}_{LV}+(Z-N)g^{(1)}_{LV}\right|^2,
\end{align}
where $p_e,E_e\approx m_\mu$ are the electron momentum and energy, respectively,
which are approximately equal to the muon mass; $Z,N$ are the
number of protons and neutrons in the nucleus, $Z_{\text{eff}}$ is the effective atomic charge and
$F_p$ is the nuclear matrix element, given in Table \ref{tab:TabNuclear} for the nuclei
we are considering in our study. Notice that the conversion rate is normalized to the muon capture rate 
$\Gamma_{\text{capt}}$. The coefficients $g^{(0,1)}_{LV}$ are given by~\cite{Kuno:1999jp,Arganda:2007jw}
\begin{subequations}
  \begin{align}
    g^{(0)}_{LV}&=\frac{1}{2}\sum_{q=u,d}\left(G_V^{(q,p)}g_{LVq}+G_V^{(q,n)}g_{LVq}\right),\\
    g^{(1)}_{LV}&=\frac{1}{2}\sum_{q=u,d}\left(G_V^{(q,p)}g_{LVq}-G_V^{(q,n)}g_{LVq}\right).
  \end{align}
\end{subequations}
We stress that we consider only vector couplings, since only the $\gamma$-penguins are relevant for this process. Also, we should note that only 
valence quarks will be relevant for our purpose, because the sea quarks, like the 
strange quark, interact effectively only through the scalar part \cite{Arganda:2007jw}. 
The coupling $g_{LVq}$ is given by 
\begin{align}
  g_{LVq}&=\frac{\sqrt{2}}{G_F}e^2Q_q\left(\frac{A_{ND}}{q^2}-A_D\right). 
\end{align}
where $Q_q$ is the quark charge.
%
%
\begin{table}[tb]
\centering
\begin{tabular}{|c|c|c|}\hline
Nucleus  & Present Bound & Future Sensitivity\\ \hline\hline
Al & $-$ & $10^{-15}-10^{-18}$ \cite{Kuno:2013mha}\\ 
Ti & 4.3$\,\times\,10^{-12}$ \cite{Dohmen:1993mp}& $\sim 10^{-18}$ \cite{Alekou:2013eta}\\ 
Au & 7$\,\times\,10^{-13}$ \cite{Bertl:2006up}& $-$\\ \hline
\end{tabular}
\caption{Present bound and future sensitivity on the $\mu-e$ nuclear conversion rate~\cite{Abada:2014cca}.
\label{tab:TabBounds}}
\end{table}
For completeness, we quote here the values of the coefficients $G_V^{(q,p)}$ \cite{Arganda:2007jw}:
\begin{align}
   G_V^{(u,p)}=G_V^{(d,n)}=2, ~~~~~ G_V^{(u,n)}=G_V^{(d,p)}=1.
\end{align}
From the present bounds on the $\mu-e$ conversion rate, Table~\ref{tab:TabBounds}, we get 
$\rho\lesssim 30$ $\mathrm{eV}^{-2}$ for titanium (Ti), and 
$\rho\lesssim 13.5$ $\mathrm{eV}^{-2}$ for gold (Au), while from the future expected sensitivities 
in aluminium (Al) and titanium we get $\rho\lesssim 0.020$ $\mathrm{eV}^{-2}$ and $\rho\lesssim 0.015$ $\mathrm{eV}^{-2}$, 
respectively. We stress that $\mu-e$ conversion in nuclei will be, in the future, the
most sensitive process to probe $\nu$2HDM, if the experiments reach the announced sensitivities. 
In Figure~\ref{fig:limits} we see how $\mu-e$ conversion sets limits on the $(m_{H^\pm}, v_2)$ 
plane using the present results (left panel) and the forecast sensitivity (right panel). 

\subsubsection{$Z\to \ell_\alpha \ell_\beta$}
%
\begin{figure}[tb]
\includegraphics[width=1.02\textwidth]{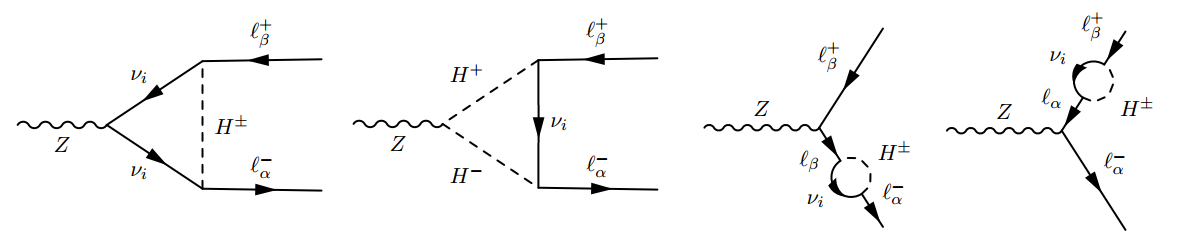} 
\vspace{-0.5cm} \caption{\label{fig:lfvz} 
Penguin and self-energy diagrams contributing to the Z LFV decays.}
\end{figure}
Besides their impact on charged lepton decays, neutrinophilic scalars 
can give rise to LFV $Z$ boson decays. 
In our framework, the additional one-loop diagrams contributing to
this process are shown in Fig.~\ref{fig:lfvz}, where $\ell_\alpha$
and $\ell_\beta$ are charged leptons of different flavors. The
effective Hamiltonian can be written as
\begin{equation}
\mathcal{H}_{\mathrm{eff}} =   C_V \bar{\ell}_\alpha \gamma^\mu P_L \ell_\beta Z_\mu \, +\mathrm{h.c.}
\end{equation}
Neglecting fermion masses, the Wilson coefficient $C_V$ is given by 
\begin{align}
C_V =\frac{1}{64\pi^2} \frac{ g\cos (2\theta_W)}{\cos \theta_W}\frac{\langle m_{\alpha\beta}^2\rangle}{v_2^2} \Bigg{\lbrace} &4 \left( \frac{2}{x_Z}-1\right) \left(\frac{4}{x_Z}-1\right)^{1/2}\arctan\left[\left(\frac{4}{x_Z}-1\right)^{-1/2} \right]\\
&-\frac{16}{x_Z^2}\arctan^2\left[\left(\frac{4}{x_Z}-1\right)^{-1/2}\right]+\left(5-\frac{4}{x_Z}\right)\Bigg{\rbrace},\nonumber
\end{align}
with $x_Z=m_Z^2/m^2_{H^\pm}<1$. In the $x_Z \ll 1$ limit this expression can be
further simplified to
\begin{equation}
C_V = \frac{x_{Z}}{288 \pi^2}\frac{ g\cos (2\theta_W)}{\cos \theta_W}\frac{\langle m_{\alpha\beta}^2\rangle}{v_2^2} +\mathcal{O}(x_Z^2).
\end{equation} 
The Wilson coefficient $C_V$ directly enters in the expression for the $Z \to \ell_\alpha \ell_\beta$ decay width, which in the limit of vanishing lepton masses is given by
\begin{equation}
\Gamma(Z\to \ell^\pm_\alpha \ell^\mp_\beta) = \dfrac{m_Z}{12 \pi \Gamma_Z}  |C_V|^2,
\end{equation}
where $m_Z$ and $\Gamma_Z$ are the $Z$ mass and total decay width and 
we took into account that $\Gamma(Z\to \ell^\pm_\alpha \ell^\mp_\beta)=\Gamma(Z\to \ell^-_\alpha \ell^+_\beta)+\Gamma(Z\to \ell^+_\alpha \ell^-_\beta)$.

On the experimental side, the most constraining bound comes from the
ATLAS upper limit $\mathrm{BR}(Z\to e^\pm\mu^\mp)<7.5 \times 10^{-7}$
\cite{Aad:2014bca}. The channels with a $\tau$ lepton in the final state
were only studied at LEP and have weaker experimental limits,
$\mathrm{BR}(Z\to e^\pm\tau^\mp)<9.8 \times 10^{-6} $ and
$\mathrm{BR}(Z\to \mu^\pm\tau^\mp)<1.2 \times 10^{-5} $
\cite{Abreu:1996mj}. 

Using ATLAS current limit on  $Z\to e^\pm\mu^\mp$ we get an upper bound 
$\rho\lesssim 3.5 \times 10^{3}$ eV$^{-2}$, much weaker than any of 
the bounds presented so far. Even considering the expected sensitivity at a future electron-positron 
collider operating at the Z pole (TLEP), $\mathrm{BR}(Z\to e^\pm\mu^\mp)\sim 10^{-13}$~\cite{Abada:2014cca}, 
the situation is not going to improve much. Instead, if we consider the current bounds on 
the parameter $\rho$ coming from $\mu\to e \gamma$, then we can predict 
$\mathrm{BR}(Z\to e^\pm \mu^\mp)\lesssim 10^{-17}$ and $\mathrm{BR}(Z\to \mu^\pm \tau^\mp)\lesssim 10^{-16}$. 

\subsubsection{$h\to \ell_\alpha \ell_\beta$}

In this section we briefly discuss the LFV process $h\to \ell_\alpha \ell_\beta$,
where $h$ denotes the SM-like Higgs. The effective Hamiltonian
describing this decay can be written as
\begin{equation}
\mathcal{H_\mathrm{eff}} = C_L \bar{\ell}_\alpha P_L \ell_\beta h \,+ \mathrm{h.c.}
\end{equation}
Assuming $m_\beta/m_\alpha\ll 1$, the Wilson coefficient $C_L$ reads
\begin{align}
C_L = -\frac{1}{8\pi^2}\frac{\langle m_{\alpha\beta}^2\rangle}{v_2^2} \frac{m_\alpha g_{hH^+ H^-}}{m_{H^\pm}^2} \frac{1}{y_H} \Bigg\lbrace & 1-2\left( \frac{4}{y_{H}}-1\right)^{1/2}\arctan\left[\left(\frac{4}{y_H}-1\right)^{-1/2}\right]\\&+\frac{4}{y_H}\arctan^2\left[\left(\frac{4}{y_H}-1\right)^{-1/2}\right]\Bigg\rbrace \, ,
\end{align}
with $y_H=m_h^2/m_{H^\pm}^2$. The coupling $g_{h H^+ H^-}$ is defined as the trilinear coupling
$h H^+ H^-$ and depends on the particular realization of the scalar
sector. In the asymptotic limit,
\begin{equation}
C_L= -\frac{1}{32\pi^2}\frac{\langle m_{\alpha\beta}^2\rangle}{v_2^2} \frac{m_\alpha g_{hH^+ H^-}}{m_{H^\pm}^2} \left[1+\frac{y_H}{9} +\mathcal{O}(y_H^2)\right].
\end{equation}
The decay rate $\Gamma(h\to \ell_\alpha^\pm \ell_\beta^\mp)=\Gamma(h\to \ell_\alpha^+ \ell_\beta^-)+\Gamma(h\to \ell_\alpha^- \ell_\beta^+)$ reads
\begin{equation}
\Gamma(h\to \ell_\alpha^\pm \ell_\beta^\mp)=\frac{(m_h^2-m_\alpha^2)^{2
}}{8\, \pi \, m_h^3}|C_L|^2.
\end{equation}
\begin{figure}[tb]
  \begin{center}
    \includegraphics[width=0.6\textwidth]{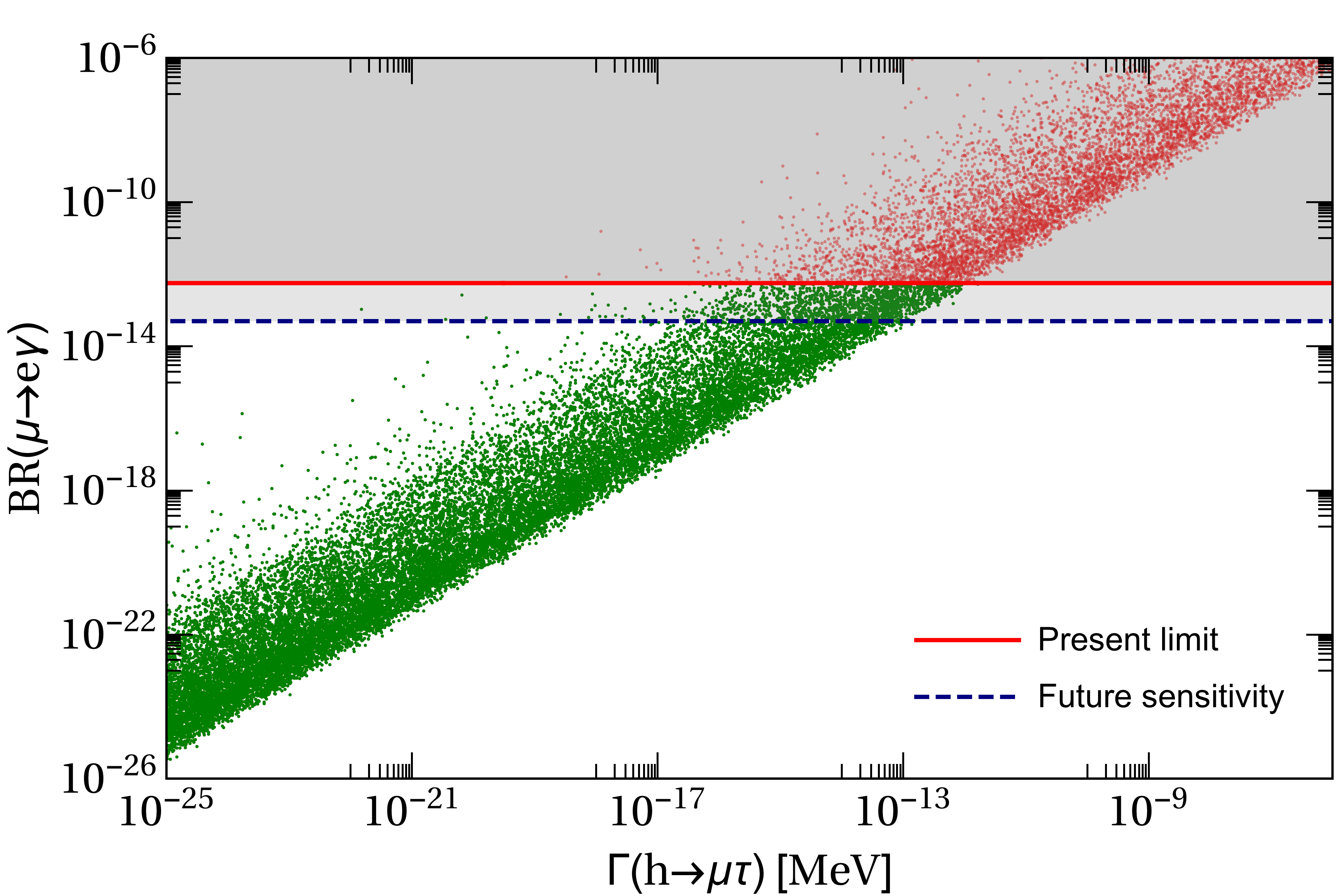}
  \end{center}
  \caption{\label{fig:higgs} $\Gamma(h \to \mu \tau)$ as a function of $\mathrm{BR}(\mu \to e \gamma)$ for the allowed parameter space of the $\nu$2HDM of Ref.~\cite{Davidson:2009ha}. The gray regions correspond to the current exclusion by MEG-2~\cite{Adam:2013mnn} (red line) and to the future expected sensitivity (blue, dashed line).}
\end{figure}
In order to make predictions for LFV Higgs decays, one needs to
consider a specific realization of the scalar potential and scan over
the parameter space allowed by theoretical and phenomenological
constraints. Here, we consider the model proposed in
Ref.~\cite{Davidson:2009ha} which, as discussed in~\cite{Machado:2015sha} is the only minimal realization of
the $\nu$2HDM still consistent with electroweak precision measurements.
Scanning over the allowed parameter space of this model (see~\cite{Machado:2015sha} for details), we obtain  a
prediction for $\Gamma(h \to \mu \tau)$, the largest LFV Higgs decay, as a function of 
$\mathrm{BR}(\mu \to e \gamma)$, as shown in Fig.~\ref{fig:higgs}.
We see that due to the stringent limit imposed by MEG-2, 
$\Gamma(h \to \mu \tau)$ cannot exceed $10^{-9}$ MeV, so this model 
cannot possibly explain a branching ratio as high as  
$\mathrm{BR}(H \to \mu \tau)= (0.84^{+0.39}_{-0.37})$ \% as
measured by CMS~\cite{Khachatryan:2015kon}.
Unfortunately, such a small branching ratio is also completely out of the reach of the LHC or even 
of any foreseen future Higgs precision experiment.

\section{Conclusions}
\label{sec:conc}

We have focused our study on the neutrinophilic two-Higgs-doublets model scenario, $\nu$2HDM, in which a neutrinophilic 
Higgs doublet is responsible for neutrino masses through a tiny vev $v_2 \lesssim 1$ eV 
and in which neutrinos are Dirac particles. Interestingly, in this scenario indirect limits 
are much more effective in constraining the parameter space than direct collider searches. 
This is due to the fact that the new scalars basically only couple to leptons and gauge bosons, 
in such a way that only the quite weak direct limit from LEP applies, $m_{H^\pm} \gtrsim 80$ GeV~\cite{Abbiendi:2013hk}.

Indirect limits can come either from lepton flavor conserving or from lepton flavor changing 
processes. The important point is that such flavor effects are controlled by the effective 
neutrino mass $\langle m_{\alpha \beta}^2 \rangle$, defined below Eq.~(\ref{eq:ratiog}), so 
that they can be well predicted now that, thanks to the measurement of the last mixing angle 
$\theta_{13}$, we are entering in the era of precision neutrino physics. This allows to put 
stringent limits on two of the unknown parameters of the $\nu$2HDM, namely the mass of the neutrinophilic
 charged Higgs boson, $m_{H^\pm}$, and the vev of the neutrinophilic doublet, $v_2$. Let us stress that, 
although the neutrinophilic charged Higgs boson modifies the neutrino propagation in matter, 
this modification affects only the right-handed neutrinos, which are not produced with a relevant rate in the sun.

In this paper we have investigated limits coming from $\mu \to e \gamma$, $\tau \to \mu (e) \gamma$, 
$\mu \to 3 e$, $\mu \to e$ in nuclei, lepton flavor violating Z and Higgs decays. Other limits, like the one coming from Big Bang Nucleosynthesis, turn out to be weaker (see Appendix~\ref{appendix}). Our main results are summarized in Fig.~\ref{fig:limits}. On the left panel we show the current bounds, which are dominated by $\mu \to e \gamma$ for $v_2 \gtrsim 0.01$ eV and by $\mu \to eee$ for $v_2 \lesssim 0.01$ eV. For example, in the region dominated by $\mu \to e \gamma$, we get a lower bound of $m_{H^\pm} \gtrsim 250 \, (2500)$ GeV for $v_2 \lesssim 1 \, (0.1)$ eV at 90\% CL, while in the region dominated by $\mu \to 3e$ the lower bound on the charged Higgs boson mass is worse than $(30-40)$ TeV. Since the philosophy of the $\nu$2HDM is to allow for ${\cal O}(1)$ Yukawa couplings in the neutrino sector, we do not expect the $v_2 \ll 0.01$ eV region to be particularly relevant.
In the right panel of Fig.~\ref{fig:limits} we instead show the future sensitivity, in which the limits could be largely dominated by $\mu-e$ conversion in nuclei. In this case, if nothing is observed, we expect the lower bound for $v_2 = 1$ eV to get as stringent as $m_{H^\pm} \gtrsim 2$ TeV, and the one for $v_2 =0.1$ eV to become $m_{H^\pm} \gtrsim 20$ TeV. 

Let us conclude mentioning that this model predicts lepton flavor violating Higgs decays, which is an interesting possibility in light of the recently CMS observation of an excess in the $h \to \mu \tau$ channel~\cite{Khachatryan:2015kon}. Unfortunately, the region compatible with the observed value for the branching ratio is already excluded by the $\mu \to e \gamma$ limit, Fig.~\ref{fig:higgs}, in such a way that such an excess cannot be observed in the $\nu$2HDM framework. Similarly, $\mathrm{BR}(Z\to \ell_\alpha \ell_\beta)$, with $\ell_\alpha\neq \ell_\beta$, is constrained to be $\lesssim 10^{-16}$ by the experimental limit on $\mu\to e \gamma$, making this observable beyond the reach of the current experiments.

\acknowledgments 

We thank Takashi Toma for useful correspondence. This work was supported by Funda\c{c}\~ao de Amparo \`a
Pesquisa do Estado de S\~ao Paulo (FAPESP) and Conselho Nacional de
Ci\^encia e Tecnologia (CNPq). We also acknowledge partial support from 
the European Union FP7 ITN INVISIBLES (Marie Curie Actions, 
PITN-GA-2011-289442).


\appendix
\section{Limits from Big Bang Nucleosynthesis}\label{appendix}

Let us now discuss the limits on the $\nu$2HDM coming from Big Bang Nucleosynthesis. The standard definition for the number of relativistic degrees of freedom $N_{eff}$, and its expression in the $\nu$2HDM, are given by
\begin{equation}\label{eq:NeffNu2HDM}
 \rho = N_{eff} \frac{7}{8} \left(\frac{4}{11}\right)^{4/3} \rho_\gamma , ~~~~~~~~N_{eff} = \left(\frac{11}{4}\right)^{4/3} 3 \left[ \left(\frac{T_{\nu_L}}{T_\gamma} \right)^4 +  \left(\frac{T_{\nu_R}}{T_\gamma} \right)^4 \right] \, ,
\end{equation}
to be compared with the result of the Planck collaboration, $N_{eff} = 3.15 \pm 0.23$~\cite{Ade:2015xua}. In order for the experimental bound to be satisfied, we must require $T_{\nu_R} \ll T_{\nu_L}$, {\it i.e.} the RH neutrinos must decouple from the thermal bath well before Big Bang Nucleosynthesis. We can estimate $T_{\nu_R}$ imposing total entropy conservation, $g_{*S}(T) a^3 T^3$ = const. We get
\begin{equation}
 \left( \frac{T_{\nu_R}}{T_\gamma} \right)^3 = \frac{172}{11 (4 g_{*S}(T_{\nu_R,d}) -21)} \, ,
\end{equation}
with $g_{*S}(T_{\nu_R,d})$ the number of relativistic degrees of freedom (in entropy) at the temperature at which the RH neutrinos decouple from the thermal bath. Following the thermal evolution of the universe backwards in time, and adding the RH neutrinos to the SM relativistic degrees of freedom, we have
\begin{equation}
 \begin{array}{lcl}
  m_\mu < T_{\nu_R,d} < m_\pi & ~~~&  g_{*S} = 39/2 \, , \\
  m_\pi < T_{\nu_R,d} < T_{quark-hadron} & ~~~&  g_{*S} = 45/2 \, ,\\
  T_{quark-hadron} < T_{\nu_R,d} < m_c & ~~~&   g_{*S} = 67 \, .\\
 \end{array}
\end{equation}
Using this result in Eq.~(\ref{eq:NeffNu2HDM}), we find $T_{\nu_R,d} > T_{quark-hadron} \sim 300$ MeV. The RH neutrinos decoupling temperature can be estimated using Eq.~(\ref{eq:Leff}) and comparing with $T_{\nu_L,d} \simeq 1$ MeV:
\begin{equation}
 \left( \frac{T_{\nu_R,d} }{T_{\nu_L,d} } \right)^3 \simeq \frac{1}{\rho^2 |\langle m_{\alpha\beta} \rangle |^2 |\langle m_{\rho\sigma} \rangle |^2} \gtrsim \left( \frac{300~{\rm MeV} }{1~{\rm MeV} } \right)^3 \, ,
\end{equation}
which can be used to extract an upper bound on $\rho$. Scanning over the netrino parameters at $2\sigma$, we find $\rho \lesssim 300~{\rm eV}^{-2}$, which is weaker than the bounds coming from flavor physics we have presented.

\end{document}